\documentclass[aps,pre,reprint,groupedaddress,nofootinbib]{revtex4-2}

\usepackage{amsmath}
\usepackage{amssymb}
\usepackage{graphicx}
\usepackage{bm}
\usepackage{xcolor}
\usepackage{siunitx}
\usepackage{tikz}
\usetikzlibrary{arrows.meta,positioning,calc,decorations.pathmorphing,
                decorations.markings,shapes.geometric,fit,backgrounds,shadows}

\newcommand{\vect}[1]{\bm{#1}}

\begin{document}

\title{A Bonded-Particle Discrete-Element Model with Self-Gravity for the Collision and Reaccumulation of Rubble-Pile Asteroids: Formulation and Numerical Protocol}

\author{Yohann Trivino}
\affiliation{Independent research}

\date{\today}

\begin{abstract}
I present a discrete-element approach developed to study the collisional and gravitational evolution of small self-gravitating granular aggregates. Each body is represented as a composite of cohesively bonded spherical grains, so that fragmentation and reaccumulation emerge from the microscopic competition between short-range dissipative contact forces, an irreversible bond-breaking criterion, and pairwise Newtonian self-gravity summed over every particle pair, including those internal to a single body. A numerical experiment is described: a random-sequential packing procedure that generates two bonded spherical aggregates, a controlled two-body encounter with tunable approach speed and impact parameter, and a systematic parameter sweep designed to map the conditions under which two bodies launched from rest under their mutual gravity alone either merge into a single remnant or fail to settle after a low-speed encounter. Of the 125 simulated collisions, 56.8\% result in the formation of a single residue. The cohesive network remains intact for $\Pi_\sigma\lesssim0.036$, is partially damaged with dispersion determined by the contact damping ratio, and breaks completely for $\Pi_\sigma\approx3.59$.

\end{abstract}

\maketitle

\section{Introduction}
\label{sec:intro}

Most asteroids smaller than a few tens of kilometers are not monolithic rocks. Their low measured bulk densities, fast rotation-fission events, binary and contact-binary morphologies, and the survival of loosely consolidated shapes such as (25143) Itokawa \cite{demura2006pole} and (162173) Ryugu \cite{tatsumi2020global} are collectively interpreted as evidence that these bodies can be treated as self-gravitating agglomerations of smaller fragments held together by a combination of gravity, geometric interlocking, and comparatively weak inter-grain cohesion, rather than by the tensile strength of a solid rock matrix \cite{walsh2018rubble, richardson2009numerical}. Understanding how such aggregates respond to a collision (whether they shatter, partially disrupt, or reaccumulate into a new equilibrium shape) is important for models of the collisional evolution of the asteroid belt, or the interpretation of crater records and satellite formation around asteroids, and, more recently, to the assessment of kinetic-impactor deflection strategies, as demonstrated by the \cite{cheng2023momentum, rivkin2021double}.

Because the relevant length scale spans from centimeter to meter-sized constituent grains up to kilometer-sized parent bodies, and because the response is intrinsically discontinuous, continuum codes are not the natural tool for the late-time, low-speed regime of the problem. The discrete-element method (DEM), introduced for granular assemblies of rigid particles \cite{Cundall1979}, provides a microscopic, particle-based description in which macroscopic strength, fragmentation, and reaccumulation are emergent properties of a small number of grain-scale contact and cohesion parameters. Soft-DEM codes coupled to self-gravity have been used extensively to study the dynamics of non-cohesive and weakly cohesive granular aggregates under their own gravity, from rubble-pile reshaping to low-speed collisions \cite{sanchez2011simulating, schwartz2012implementation, trivino2025gravitational}. Cohesion between grains is commonly introduced through a bonded-particle model in which pairs of grains are linked by an elastic link that transmits normal and shear loads until a critical stress is exceeded, at which point the bond fails irreversibly \cite{potyondy2004bonded, schwartz2013numerically, zhang2018rotational}.

The present paper documents a compact, from-scratch DEM model built to explore this dynamic in a self-contained and reproducible setting: two finite composite bodies, each a random packing of a few hundred or a few thousand bonded spherical grains, are brought into contact either through an imposed approach velocity or through free-fall under their mutual self-gravity alone, and the subsequent evolution of the bond network and of the two-body orbital state is measured. The model deliberately couples three physical ingredients that are each individually standard in the granular-mechanics and planetary-science literature: dissipative soft-sphere contact \cite{Cundall1979, silbert2001granular, schwartz2012implementation}, irreversible cohesive bonding \cite{potyondy2004bonded, schwartz2013numerically, sanchez2012dem}, and direct $N$-body self-gravity \cite{richardson2000direct, leinhardt2000direct, richardson2009numerical}. But these features are comparatively rarely exercised together in a single study which is the choice made here.

The paper is organized as follows. Section \ref{sec:model} gives the complete formulation of the model: the particle representation, the non-bonded contact law, the cohesive bonded-particle model and its failure criterion, the direct-summation self-gravity, the Verlet neighbor-list acceleration and the time integrator. Section \ref{sec:experiment} describes the protocol of simulation built for the study: the stochastic generation of a bonded rubble pile, the two-body encounter and the systematic parameter sweep used to probe the conditions for gravitational reaccumulation.

The model is first validated against the exact solution for two-body free fall up to the moment of contact. I then show that the response of the link network is organized around a single dimensionless impact stress number, which distinguishes between the "undamaged", "partially damaged", and "completely fragmented" regimes. And the number of low-speed rebounds required for the two bodies to merge into a single fragment is characterized as a function of the damping ratio at contact.

\section{Model}
\label{sec:model}

\subsection{Composite bodies and particle representation}
\label{sec:model:particles}

The fundamental degree of freedom of the model is the rigid sphere. Particle $k$ is fully specified by a position $\vect{x}_k$, a velocity $\vect{v}_k$, a radius $r_k$, and a density $\rho_k$, from which the mass follows as $m_k = \tfrac{4}{3}\pi \rho_k r_k^3$. Each particle additionally carries an integer label $b_k$ identifying the composite body to which it belongs. Two particles are only ever candidates for a permanent cohesive bond (Sec \ref{sec:model:bond}) if they share the same $b_k$, but they interact through contact (Sec \ref{sec:model:contact}) and through gravity (Sec \ref{sec:model:gravity}) irrespective of body membership. Here, a composite asteroid is nothing more than a set of particles sharing a label and a network of cohesive bonds. No rigid-body or continuum degrees of freedom are introduced at the composite level. The composite's translational and any effective rotational or elastic, behavior is entirely emergent from the sum of its grains' interactions. Rotational degrees of freedom of the individual grains themselves are not modeled (only translational motion is integrated) so that contact and bond moments are absent from the formulation. I address to this limitation in Sec \ref{sec:model:bond}.

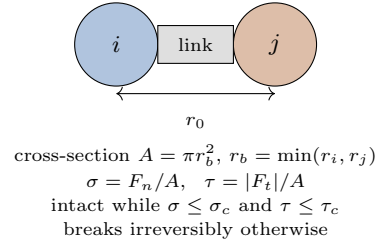
\begin{figure}[tbh]
\centering
\begin{tikzpicture}[
    scale=1.0,
    every node/.style={font=\small},
    ]
   \node[font=\bfseries\small] at (0.0,3.35) {(a) non-bonded contact};
   \begin{scope}
     \definecolor{grainA}{RGB}{176,196,222}
     \definecolor{grainB}{RGB}{221,190,169}
     \fill[grainA] (-0.55,2.2) circle (0.62);
     \fill[grainB] (0.55,2.2) circle (0.62);
     \draw (-0.55,2.2) circle (0.62);
     \draw (0.55,2.2) circle (0.62);
     \node at (-0.55,2.2) {$i$};
     \node at (0.55,2.2) {$j$};
     \fill[gray!55,opacity=0.6] (-0.02,2.2) ellipse (0.14 and 0.5);
     \draw[gray!70] (-0.02,2.2) ellipse (0.14 and 0.5);
     \draw[-{Latex[length=1.6mm]}] (-1.55,1.15) -- (-0.9,1.55);
     \draw[thick] (-1.7,0.75) -- (-1.7,1.6) node[above]{};
     \draw[decorate,decoration={coil,aspect=0.6,segment length=2.2mm,amplitude=1.4mm}]
        (-1.7,0.85) -- (-0.9,1.5) node[midway,left=1mm]{$k_n$};
     \draw[thick] (-1.4,0.85) rectangle (-1.15,1.15);
     \draw (-1.4,1.0) -- (-1.15,1.0);
     \node[align=left] at (-0.95,0.55) {\footnotesize $\gamma_n$};
     \draw[decorate,decoration={coil,aspect=0.6,segment length=2mm,amplitude=1.1mm}]
        (0.9,0.6) -- (1.7,0.6) node[midway,above=1mm]{\footnotesize $k_t$};
     \draw[thick] (1.7,0.45) rectangle (2.05,0.75);
     \node at (2.35,0.6) {\footnotesize $\mu$};
     \draw[-{Latex[length=1.4mm]}] (2.35,0.75) -- (2.35,1.05);
     \node[align=center,font=\scriptsize] at (0.0,-0.05)
        {$F_n=k_n\delta_n-\gamma_n v_n\ (\ge 0)$\\[1pt]
         $|F_t|\le \mu F_n$, no tension};
   \end{scope}

   \draw[gray!60] (-2.3,-0.55) -- (2.3,-0.55);

   \begin{scope}[yshift=-4.25cm]
     \node[font=\bfseries\small] at (0.0,3.35) {(b) cohesive bond (intact)};
     \definecolor{grainA}{RGB}{176,196,222}
     \definecolor{grainB}{RGB}{221,190,169}
     \fill[grainA] (-1.05,2.2) circle (0.55);
     \fill[grainB] (1.05,2.2) circle (0.55);
     \draw (-1.05,2.2) circle (0.55);
     \draw (1.05,2.2) circle (0.55);
     \node at (-1.05,2.2) {$i$};
     \node at (1.05,2.2) {$j$};
     \fill[gray!25] (-0.5,1.95) rectangle (0.5,2.45);
     \draw (-0.5,1.95) rectangle (0.5,2.45);
     \node[font=\scriptsize] at (0,2.2) {link};
     \draw[<->] (-1.05,1.55) -- (1.05,1.55) node[midway,below=2mm,font=\scriptsize]{$r_0$};
     \node[align=center,font=\scriptsize,anchor=north] at (0.0,1.0)
        {cross-section $A=\pi r_b^2$, $r_b=\min(r_i,r_j)$\\[2pt]
         $\sigma=F_n/A,\ \ \tau=|F_t|/A$\\[1pt]
         intact while $\sigma\le\sigma_c$ and $\tau\le\tau_c$\\[1pt]
         breaks irreversibly otherwise};
   \end{scope}
\end{tikzpicture}
\caption{Rheological content of the two short-range interaction channels. (a) Two unbonded, overlapping grains interact through a soft-sphere contact: a linear spring-dashpot along the normal direction, active only in compression ($\delta_n>0$), and a memory-bearing tangential spring capped by Coulomb friction. (b) Two grains of the same parent body that were close enough at generation time are permanently linked by an elastic cohesive bond of cross-section $A=\pi r_b^2$, $r_b=\min(r_i,r_j)$, until the transmitted normal or shear stress exceeds a critical threshold, at which point the bond fails once and for all.}
\label{fig:rheology}
\end{figure}

\subsection{Non-bonded contact interaction}
\label{sec:model:contact}

Two spheres $i,j$ that overlap, $\delta_n \equiv r_i+r_j-|\vect{x}_j-\vect{x}_i| > 0$, interact through a Cundall-Strack linear model~\cite{Cundall1979}, illustrated in Fig \ref{fig:rheology}(a), irrespective of whether they belong to the same or to different composite bodies: once a cohesive bond between two grains has failed (Sec \ref{sec:model:bond}), this is precisely the interaction that takes over when the two grains remain in geometric contact. Let $\vect{n}_{ij}=(\vect{x}_j-\vect{x}_i)/|\vect{x}_j-\vect{x}_i|$ be the unit normal, $\vect{v}_{ij}=\vect{v}_j-\vect{v}_i$ the relative velocity, and $v_n=\vect{v}_{ij}\cdot\vect{n}_{ij}$, $\vect{v}_t = \vect{v}_{ij} - v_n\vect{n}_{ij}$ its normal and tangential components. The normal contact force is
\begin{equation}
F_n = \max\!\bigl(0,\ k_n\,\delta_n - \gamma_n v_n\bigr),
\label{eq:fn_contact}
\end{equation}
with $k_n$ a normal contact stiffness and $\gamma_n$ a normal damping coefficient. The clipping at zero enforces that contact is purely repulsive (i.e. two grains that are not bonded exert no adhesive force on one another once they separate).

The tangential response uses an incremental elastic-predictor, Coulomb-corrector scheme with memory. A per-contact tangential elastic displacement $\vect{u}_t$ is carried from one time step to the next, re-projected onto the current tangential plane to remove any drift along the (possibly rotating) normal direction, and advanced according to $\vect{u}_t \leftarrow (\vect{u}_t - (\vect{u}_t\!\cdot\!\vect{n}_{ij})\vect{n}_{ij}) + \vect{v}_t\,\Delta t$.
A trial tangential force,
\begin{equation}
\vect{F}_t^{\text{trial}} = -k_t\,\vect{u}_t - \gamma_t\,\vect{v}_t ,
\end{equation}
is then clipped to satisfy the Coulomb inequality $|\vect{F}_t|\le \mu F_n$: if $|\vect{F}_t^{\text{trial}}| > \mu F_n$ the contact is sliding and the transmitted force is rescaled, $\vect{F}_t=\vect{F}_t^{\text{trial}}\,\mu F_n/|\vect{F}_t^{\text{trial}}|$, with the stored displacement $\vect{u}_t$ simultaneously corrected so that re-inserting it through the elastic law above reproduces the clipped force exactly, keeping the memory variable consistent with what was actually applied. The pair $(\vect{F}_n,\vect{F}_t)$ act along $\vect{n}_{ij}$ and within the tangential plane respectively and are applied as an action-reaction pair, $\vect{F}_{i\to j}=-\vect{F}_{j\to i}$. Contacts that stop to overlap have their tangential memory discarded. Because this short-range interaction must in principle be evaluated for every pair of grains at every time step, its cost is controlled by restricting the candidate pairs to a Verlet neighbor list (Sec \ref{sec:model:verlet}) rather than by rescanning all $O(N^2)$ pairs.

\subsection{Cohesive bonded-particle model}
\label{sec:model:bond}

Cohesion between grains of a common parent body is introduced through a bonded-particle model sketched in Fig \ref{fig:rheology}(b). At sample-generation time (detailed in Sec \ref{sec:experiment:generator}), pairs of grains in the same body that are sufficiently close are glued by an elastic link of cross-sectional area:
\begin{equation}
A = \pi r_b^2, \qquad r_b = \min(r_i,r_j)
\end{equation}
recording the pair's initial center-to-center distance $r_0$ as its natural, unstressed length. Unlike the contact law of Sec \ref{sec:model:contact}, an intact bond transmits both tensile and compressive normal load: with $\delta_n \equiv r-r_0$ the elongation at the current separation $r=|\vect{x}_j-\vect{x}_i|$,
\begin{equation}
F_n^{\text{bond}} = k_n^{\text{bond}}\,\delta_n ,
\end{equation}
positive $\delta_n$ (stretching) pulling the two grains together and negative $\delta_n$ (compression) pushing them apart. The bond behaves as a simple linear spring around $r_0$, with no rectification at $\delta_n=0$. The shear response accumulates a tangential elastic displacement $\vect{u}_t^{\text{bond}}$ exactly as in Sec \ref{sec:model:contact}, giving a shear force $\vect{F}_t^{\text{bond}}=k_t^{\text{bond}}\,\vect{u}_t^{\text{bond}}$ with no Coulomb cap. So while intact, a bond is a permanent elastic link, not a frictional contact.

The bond carries a normal and a shear stress,
\begin{equation}
\sigma = \frac{F_n^{\text{bond}}}{A}, \qquad
\tau = \frac{|\vect{F}_t^{\text{bond}}|}{A},
\label{eq:bond_stress}
\end{equation}
evaluated at every time step and compared against critical thresholds $\sigma_c$ (tensile) and $\tau_c$ (shear). The bond is intact as long as $\sigma\le\sigma_c$ and $\tau\le\tau_c$. As soon as either threshold is exceeded, the bond is marked broken and thereafter transmits no force for the remainder of the simulation. Failure is an irreversible event and a fixed compressive threshold is used symmetrically for $\sigma<0$ since the criterion in Eq \ref{eq:bond_stress} is applied to the signed stress. Once broken, the pair of grains it used to connect reverts to ordinary, tensionless contact (Eq \ref{eq:fn_contact}) if they remain in geometric overlap. This is the microscopic mechanism by which a composite body absorbs a disturbance elastically but sheds discrete bonds and/or fragments, under a sufficiently violent one. This is a stress, rather than an energy-based, failure criterion, applied (not combined, e.g., in a Mohr-Coulomb or elliptical envelope) to the normal and shear channels. 

\subsection{Self-gravity}
\label{sec:model:gravity}

Every pair of particles in the simulation, whether or not they belong to the same composite body, attracts every other pair through Newtonian gravity, softened at short range to regularize near-coincident grains:
\begin{equation}
\vect{F}_i^{\text{grav}} = \sum_{j\ne i} \frac{G\,m_i m_j}
{\bigl(|\vect{x}_j-\vect{x}_i|^2+\epsilon^2\bigr)^{3/2}}\,
(\vect{x}_j-\vect{x}_i),
\label{eq:gravity}
\end{equation}
where $\epsilon$ is a softening length (set to zero unless stated otherwise) and $G$ is the gravitational constant. The sum runs over all particles in the simulation without regard to body membership: it is this term, evaluated intra-body, that (together with cohesion) holds a rubble pile together in the first place and, inter-body, that lets two initially separate composites attract each other and, potentially, remain gravitationally bound after a gentle encounter, whether or not any residual bond survives between them. 

Equation \ref{eq:gravity} is evaluated by direct pairwise summation. It is an unavoidable $O(N^2)$ cost per time step for a genuinely long-range interaction with no natural notion of a neighbor cutoff. The double loop over $i$ and $j$ is distributed over threads by particle index $i$, each thread privately accumulating $\vect{F}_i^{\text{grav}}$, which avoids any write conflict between threads at the cost of a factor of two in floating-point operations relative to a half loop with explicit action-reaction bookkeeping. 

\subsection{Verlet neighbor list \cite{verlet1967computer}}
\label{sec:model:verlet}

Rescanning all pairs at every time step merely to determine which grains are close enough for the short-range interactions of Secs \ref{sec:model:contact}--\ref{sec:model:bond} to matter would be wasteful, since the great majority of pairs are not in contact at any given instant. The neighbor list retains, at intervals, every pair $(i,j)$, $i<j$, whose center-to-center distance is below $r_i+r_j+s$, where $s$ pads the true contact threshold $r_i+r_j$ with a safety margin. Once built, the list remains valid as long as no single particle has moved by more than $s/2$ from the position it held at the last rebuild: two particles absent from the list can approach each other by at most $2\times(s/2)=s$ before their true separation could possibly have dropped below $r_i+r_j$, which is exactly the margin built into the cutoff.

The rebuild criterion is therefore
\begin{equation}
\max_k |\vect{x}_k - \vect{x}_k^{\text{ref}}| > s/2 ,
\end{equation}
tracked against the reference configuration $\vect{x_k^{\text{ref}}}$ recorded at the previous rebuild. Between rebuilds, the contact model of Sec \ref{sec:model:contact} only tests true overlap among the candidate pairs returned by the list, rather than over all $O(N^2)$ pairs. The list narrows the candidate set but does not itself decide contact. The $s$ parameter is a pure performance parameter (by default I use $0.3$ times the mean grain radius) trading rebuild frequency against candidate-list size, with no effect on the converged physical trajectory in the small time-step limit. As noted above, this acceleration structure is specific to the short-range channels: self-gravity is evaluated over the full particle set at every step regardless.

\subsection{Time integration and numerical parameters}
\label{sec:model:integration}

At a given instant, the total force on particle $i$ is the sum of three contributions,
\begin{equation}
\vect{F}_i = \vect{F}_i^{\text{grav}} + \vect{F}_i^{\text{contact}} + \vect{F}_i^{\text{bond}},
\end{equation}
with the restriction that a pair of grains still linked by an intact bond is excluded from the non-bonded contact sum of Sec \ref{sec:model:contact} to avoid double-counting their normal interaction. 

The equations of motion, $m_k\ddot{\vect{x}}_k=\vect{F}_k$, are integrated with a Verlet scheme  \cite{swope1982computer} at fixed time step $\Delta t$,
\begin{align}
\vect{v}_k^{n+1/2} &= \vect{v}_k^{n} + \frac{\Delta t}{2 m_k}\vect{F}_k^{n},\\
\vect{x}_k^{n+1} &= \vect{x}_k^{n} + \Delta t\,\vect{v}_k^{n+1/2}, \\
\vect{v}_k^{n+1} &= \vect{v}_k^{n+1/2} + \frac{\Delta t}{2 m_k}\vect{F}_k^{n+1},
\label{eq:vv}
\end{align}
where $\vect{F}_k^{n+1}$ is evaluated at the updated positions $\vect{x}_k^{n+1}$ using the half-step velocities $\vect{v}_k^{n+1/2}$ for every velocity-dependent term. Only two force evaluations per step are needed: one at the start-of-step configuration for the half-step kick, and one at the predicted end-of-step configuration, which simultaneously serves as the corrector force and, on steps where output is due, supplies the detailed per-contact and per-bond records, so that no additional force pass is required purely for reporting purposes.

The neighbor list of Sec \ref{sec:model:verlet} is checked, and rebuilt (if necessary) once per step, immediately before the first evaluation. The time step is chosen from the classical critical time of a linear spring-mass contact,
\begin{equation*}
\Delta t = \eta\,T_c, \qquad
T_c = 2\pi\sqrt{\frac{m_{\min}}{2 k_n}},
\label{eq:dt}
\end{equation*}
with $m_{\min}$ the mass of the lightest grain in the sample and a safety factor $\eta=1/30$.

\begin{figure*}[tbh]
\centering
\begin{tikzpicture}[
    node distance=7mm and 9mm,
    box/.style={draw,rounded corners=1.5pt,align=center,minimum height=8mm,
                fill=blue!5,inner sep=3pt,font=\small},
    dec/.style={draw,diamond,aspect=2.6,align=center,inner sep=1pt,
                fill=orange!10,font=\small},
    io/.style={draw,rounded corners=4mm,align=center,minimum height=8mm,
               fill=green!8,font=\small},
    arr/.style={-{Latex[length=2mm]}},
    every node/.style={font=\small}
    ]
  \node[io]  (start) {particles $\vect{x}^n,\vect{v}^n$};
  \node[dec, right=of start] (rebuild)
       {$\max_k|\vect{x}_k-\vect{x}_k^{\rm ref}|$\\$>s/2$?};
  \node[box, above=of rebuild] (verlet) {rebuild Verlet list\\(Sec.~\ref{sec:model:verlet})};
  \node[box, right=of rebuild] (force1) {force pass 1:\\gravity + contact + bond\\at $\vect{x}^n,\vect{v}^n$};
  \node[box, right=of force1] (kick1) {half-kick + drift\\Eq.~\eqref{eq:vv}, first two lines};
  \node[box, below=of kick1] (force2) {force pass 2:\\gravity + contact + bond\\at $\vect{x}^{n+1},\vect{v}^{n+1/2}$};
  \node[box, left=of force2] (kick2) {half-kick\\Eq.~\eqref{eq:vv}, third line};
  \node[dec, left=of kick2] (due) {output\\due?};
  \node[box, below=of due] (write) {write VTK frame,\\measure row};
  \node[io, left=of due] (endstep) {particles $\vect{x}^{n+1},\vect{v}^{n+1}$};

  \draw[arr] (start) -- (rebuild);
  \draw[arr] (rebuild) -- node[above,font=\scriptsize]{yes} (verlet);
  \draw[arr] (verlet) -| (force1);
  \draw[arr] (rebuild) -- node[below,font=\scriptsize]{no} (force1);
  \draw[arr] (force1) -- (kick1);
  \draw[arr] (kick1) -- (force2);
  \draw[arr] (force2) -- (kick2);
  \draw[arr] (kick2) -- (due);
  \draw[arr] (due) -- node[right,font=\scriptsize]{yes} (write);
  \draw[arr] (write) -- (endstep);
  \draw[arr] (due) -- node[above,font=\scriptsize]{no} (endstep);
  \draw[arr] (endstep.north) -- ++(0,0.75) -| node[near start,above,font=\scriptsize]{next step $n{+}1$} (start.north);
\end{tikzpicture}
\caption{One time step of the simulation loop. The Verlet list (Sec \ref{sec:model:verlet}) is rebuilt only when required.}
\label{fig:flowchart}
\end{figure*}
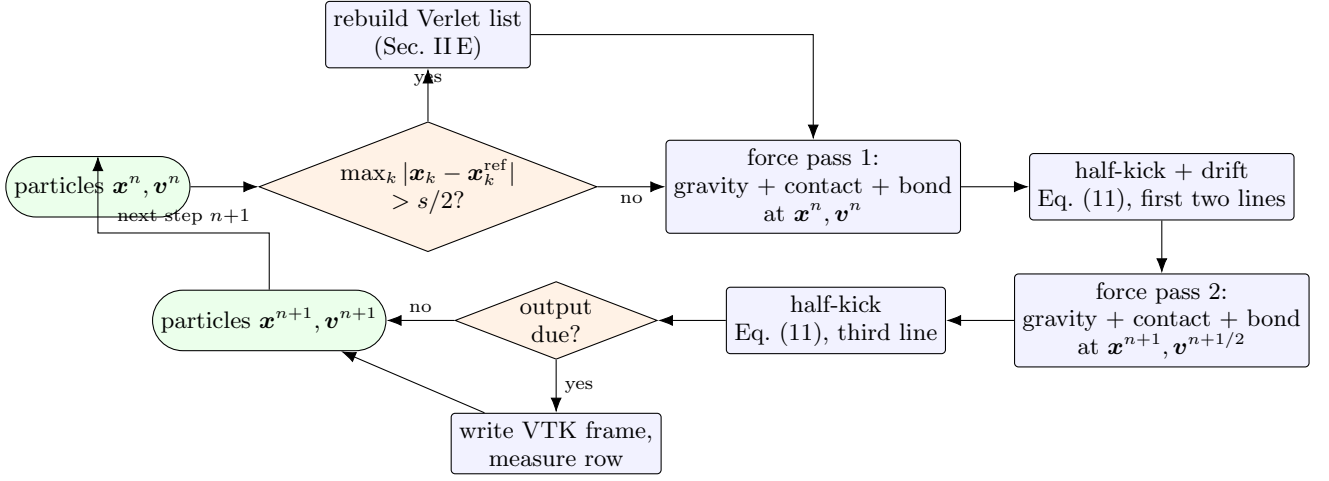

Figure \ref{fig:flowchart} summarizes the resulting per-step control flow.

Table \ref{tab:params} collects the free parameters of the model with the default value chosen. No intrinsic unit system is imposed by the code itself, and the rescaling of $G$ used in the reaccumulation sweep of Sec \ref{sec:experiment:sweep} is a property of the numerical simulation, not of the model.

\begin{table}[tbh]
\caption{Free parameters of the model (Sec.~\ref{sec:model}) and their
default values when not otherwise specified.}
\label{tab:params}
\footnotesize
\begin{ruledtabular}
\begin{tabular}{lp{4.1cm}l}
Symbol & Meaning & Default \\ \hline
$\Delta t$ & time step & Eq.~\eqref{eq:dt} \\
$k_n$ & normal contact stiffness & -- \\
$k_t$ & tangential contact stiffness & $0.8\,k_n$ \\
$\gamma_n,\gamma_t$ & normal/tangential contact damping & $0$ \\
$\mu$ & Coulomb friction coefficient & $0.5$ \\
$k_n^{\text{bond}},k_t^{\text{bond}}$ & bond stiffnesses & $k_n,\,k_t$ \\
$\sigma_c$ & critical tensile bond stress & $10^{30}\,\si{Pa}$ \\
$\tau_c$ & critical shear bond stress & $\sigma_c$ \\
$G$ & gravitational constant & $\SI{6.674e-11}{}$ \\
$\epsilon$ & gravitational softening length & $0$ \\
$s$ & Verlet skin & $0.3\,\bar r$ \\
$\rho$ & grain density (if unset per grain) & $\SI{2000}{kg/m^3}$ \\
\end{tabular}
\end{ruledtabular}
\end{table}

\section{Numerical experiment}
\label{sec:experiment}

This section describes the generator used throughout the study and the two numerical experiments built on top of it: a controlled, imposed-velocity two-body encounter, and a systematic parameter sweep for gravity-only reaccumulation.

\subsection{Generation of a pristine bonded aggregate}
\label{sec:experiment:generator}

Each composite asteroid is built independently by a random-sequential addition (RSA) packing procedure \cite{torquato2006random}. Grain radii are chosen once from a mean value, $r_k = \bar r\,(1+\xi\,(2u_k-1))$ with $u_k\sim\mathcal U(0,1)$ and $\xi$ a relative fraction and grains are then inserted one at a time at a uniformly random direction and a radially biased distance inside a sphere of radius $R$, subject to non-overlap with every previously placed grain. Numerically, I have done a criterion that if an insertion fails after a fixed number of trials, it aborts the packing at its current occupancy rather than looping indefinitely. The confining radius $R$ is set from a target volume filling fraction $\phi=0.35$ (something like a loose granular packing)
\begin{equation}
R = \bar r\,\Bigl(\frac{N}{\phi}\Bigr)^{1/3}
\end{equation}
with $N$ the number of grains per body. Once a body's grains are placed, any pair within a tolerance of first contact, $|\vect{x}_i-\vect{x}_j| \le \beta\,(r_i+r_j)$ with $\beta$ a bond-tolerance factor slightly above unity (the default value is $1.05$, absorbing the residual gaps of the RSA packing), is recorded as a permanent cohesive bond (Sec \ref{sec:model:bond}). Each asteroid is generated independently with its own random realization of this procedure (same $N$, $\bar r$, $\zeta$, $\phi$, $\beta$ for both bodies in every run reported here), then rigidly translated to a prescribed center and given a common translational velocity, without otherwise perturbing its internal configuration.

\subsection{Two-body encounter geometry}
\label{sec:experiment:encounter}

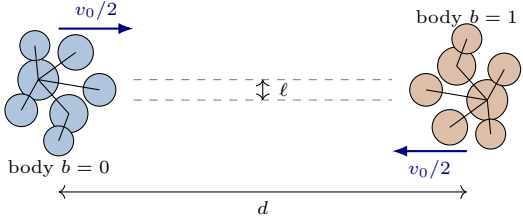
\begin{figure}[tbh]
\centering
\begin{tikzpicture}[scale=0.9,every node/.style={font=\small}]
  \definecolor{grainA}{RGB}{176,196,222}
  \definecolor{grainB}{RGB}{221,190,169}
  \begin{scope}
    \foreach \x/\y/\rr in {-3.3/0.15/0.30, -2.75/0.55/0.26, -2.85/-0.35/0.28,
                            -3.55/-0.3/0.24, -3.0/-0.75/0.22, -2.4/0.0/0.24,
                            -3.35/0.65/0.22} {
      \fill[grainA] (\x,\y) circle (\rr);
      \draw[thin] (\x,\y) circle (\rr);
    }
    \draw[thin] (-3.3,0.15)--(-2.75,0.55) (-3.3,0.15)--(-2.85,-0.35)
                (-3.3,0.15)--(-3.55,-0.3) (-3.3,0.15)--(-2.4,0.0)
                (-3.3,0.15)--(-3.35,0.65) (-2.85,-0.35)--(-3.0,-0.75);
  \end{scope}
  \begin{scope}
    \foreach \x/\y/\rr in {3.3/-0.15/0.30, 2.75/-0.55/0.26, 2.85/0.35/0.28,
                            3.55/0.3/0.24, 3.0/0.75/0.22, 2.4/0.0/0.24,
                            3.35/-0.65/0.22} {
      \fill[grainB] (\x,\y) circle (\rr);
      \draw[thin] (\x,\y) circle (\rr);
    }
    \draw[thin] (3.3,-0.15)--(2.75,-0.55) (3.3,-0.15)--(2.85,0.35)
                (3.3,-0.15)--(3.55,0.3) (3.3,-0.15)--(2.4,0.0)
                (3.3,-0.15)--(3.35,-0.65) (2.85,0.35)--(3.0,0.75);
  \end{scope}
  \node[font=\scriptsize] at (-3.0,-1.15) {body $b=0$};
  \node[font=\scriptsize] at (3.0,1.05) {body $b=1$};
  \draw[<->] (-3.0,-1.5) -- (3.0,-1.5) node[midway,below,font=\scriptsize]{$d$};
  \draw[dashed,gray] (-1.9,0.15) -- (1.9,0.15);
  \draw[dashed,gray] (-1.9,-0.15) -- (1.9,-0.15);
  \draw[<->] (0.0,0.15) -- (0.0,-0.15) node[midway,right=1mm,font=\scriptsize]{$\ell$};
  \draw[-{Latex[length=2mm]},thick,blue!50!black] (-3.0,0.9) -- (-1.9,0.9)
        node[midway,above,font=\scriptsize]{$v_0/2$};
  \draw[-{Latex[length=2mm]},thick,blue!50!black] (3.0,-0.9) -- (1.9,-0.9)
        node[midway,below,font=\scriptsize]{$v_0/2$};
\end{tikzpicture}
\caption{Two-body encounter geometry (Sec \ref{sec:experiment:encounter}). Each body is an independently packed, fully-bonded composite of $N$ grains (schematic, small $N$ shown). The two centers of mass start at separation $d$, offset by an impact parameter $\ell$, and approach at a combined relative speed $v_0$ split symmetrically between the two bodies. Setting $v_0=0$ instead reduces the setup to the pure gravitational-infall configuration used in the reaccumulation sweep of Sec \ref{sec:experiment:sweep}.}
\label{fig:encounter}
\end{figure}

The two composite bodies are placed symmetrically about the origin along the $x$ axis at a prescribed center-to-center separation $d$ (by default $6R$, comfortably outside any initial overlap) and launched toward each other at a prescribed relative approach speed $v_0$ split symmetrically between the two centers of mass, as depicted in Fig \ref{fig:encounter}. 

\subsection{Reaccumulation time scale for two bodies released from rest}
\label{sec:experiment:sweep}

A distinct, complementary numerical experiment addresses the opposite regime: two bodies released from rest, $v_0=0$, so that the only force bringing them together is their mutual self-gravity (Eq \ref{eq:gravity}, evaluated inter-body), and the question of interest is not whether a collision occurs, but whether repeated low-speed contact under self-gravity leads, within a finite simulated time, to a single merged remnant, or instead to one or more bounces that have not yet settled.

Two points of orbital mechanics fix the shape of this simulation. First, at $v_0=0$ the total angular momentum of the two-body system about their common center of mass is identically zero, $\vect{L}=\sum_b \vect{r}_b\times m_b\vect{v}_b=\vect{0}$ at $t=0$, and remains so for all time by conservation, since the mutual gravitational force is central: an initial transverse offset (the impact parameter $\ell$ of Sec \ref{sec:experiment:encounter}) therefore has no effect whatsoever on the ensuing trajectory when $v_0=0$ The impact parameter is accordingly dropped from this sweep entirely because it only becomes a physically meaningful control variable once $v_0\ne 0$, as in Sec \ref{sec:experiment:encounter}. Second, the total mechanical energy at release, $E = 0 + U_{\text{grav}}(d) < 0$, is negative for any finite starting separation $d$, so the pair is gravitationally bound in the wide sense from the outset. Since contact damping $\gamma_n$ can only remove mechanical energy, never add it, the two bodies can never escape to infinity. 

So, the physical question is not whether the two bodies remain associated (obviously... they always do) but how quickly the encounter converges to a single aggregate as opposed to bouncing without yet having merged by the end of the simulated window. This distinction motivates the choice of the three parameters actually varied in the sweep:

\begin{itemize}
\item the grain density $\rho$, which at fixed geometry sets each body's mass and hence the depth of the mutual gravitational potential well.
\item the initial separation $d$, which sets the gravitational potential energy released before first contact, and so the impact speed.
\item the normal contact damping $\gamma_n$, which sets how each individual bounce is damped (the range is in accordance with \cite{trivino2025gravitational}).
\end{itemize}

For each of the $5\times5\times5=125$ combinations drawn from five logarithmically spaced densities spanning four decades, $\rho\in\{10^{1},10^{2},10^{3},10^{4},10^{5}\}\,\si{kg/m^3}$, five logarithmically spaced initial separations spanning two decades, $d\in\{10^{1.5},10^{2},10^{2.5},10^{3},10^{3.5}\}\,\si{m}\approx \{31.6,\,100,\,316,\,1000,\,3162\}\,\si{m}$, and five logarithmically spaced contact damping coefficients spanning four decades, $\gamma_n\in\{10^{2},10^{3},10^{4},10^{5},10^{6}\}\,\si{N\!\cdot\!s/m}$ (grain number, radius, contact and bond stiffness, and the critical bond stress are held fixed across the sweep at a target $N=500$ grains per body, $\bar r=\SI{1}{m}$, $k_n=\SI{1e6}{N/m}$, $\sigma_c=\SI{1e5}{Pa}$ the generator of Sec \ref{sec:experiment:generator} is invoked with $v_0=0$ and the corresponding $(\rho,d,\gamma_n)$ triple, followed by a full run of the code of Sec \ref{sec:model} on the resulting sample, in its own, independent output subdirectory. The logarithmic spacing of $\rho$ and $\gamma_n$ is deliberate since the critical contact damping of a single grain-grain contact, $\gamma_n^{\text{crit}} = 2\sqrt{k_n m_{\text{eff}}}$ with $m_{\text{eff}}$ the reduced mass of the two grains, itself scales as $\sqrt{\rho}$, a linear grid in either variable alone would sample the physically relevant transition $\gamma_n/\gamma_n^{\text{crit}}\sim1$ only over a narrow, density-dependent corner of the grid.

Each of the $125$ grid points is generated from a single random packing realization. The reported scatter therefore convolves the physical response with the stochastic variability of the generator, which is not separately quantified here.

Two practical points complete the protocol. First, the gravitational constant used in this study, $G_{\text{sweep}}=\SI{2e-5}{}$, is deliberately amplified by six orders of magnitude above the physical value $G=\SI{6.674e-11}{}$. At the toy length and mass scale adopted here (meter-sized grains, few-tonne bodies), the physical free-fall time between the two bodies would run to simulated weeks, i.e., far beyond a tractable number of time steps at the resolution of Eq \ref{eq:dt}. 

Second, the number of integration steps allotted to each run is set from an estimate of the two-point-mass radial free-fall time (the degenerate, zero-angular-momentum limit of the Kepler problem),
\begin{equation}
t_{\text{ff}}(d) = \frac{\pi}{2\sqrt2}\,\frac{d^{3/2}}{\sqrt{G_{\text{sweep}} M_{\text{tot}}}},
\end{equation}
with $M_{\text{tot}}$ the combined mass of the two bodies at the sampled density, multiplied by a safety factor of $4$ to allow for the initial fall, the first collision, and a few subsequent bounces, and capped at a fixed maximum of $5\times10^6$ steps regardless of how large the free-fall estimate turns out to be for the most extreme corner of the parameter grid.

\section{Results}
\label{sec:results}

The random-sequential packing of Sec \ref{sec:experiment:generator}, run once at a target of $N=500$ grains per body, saturates before reaching that target: the insertion-trial cap of the packing routine is reached at $407$ and $401$. This section reports the analysis extracted from the simulations.

\subsection{Validation against the exact free-fall solution}
\label{sec:results:freefall}

Before assessing the outcome of the sweep, its self-gravity and integrator can first be checked against Sec \ref{sec:experiment:sweep}'s own analytic reference: the zero-angular-momentum Kepler solution $x(\tau)$, $\dot x(\tau)$ parameterized by Eq \ref{eq:dt}'s companion angle variable (Sec \ref{sec:experiment:sweep}), universal and free of any run parameter once separation and time are rescaled by $d_0$ and $t_{\text{ff}}(d_0)$ respectively. 

\begin{figure}[tbh]
\centering
\includegraphics[width=\columnwidth]{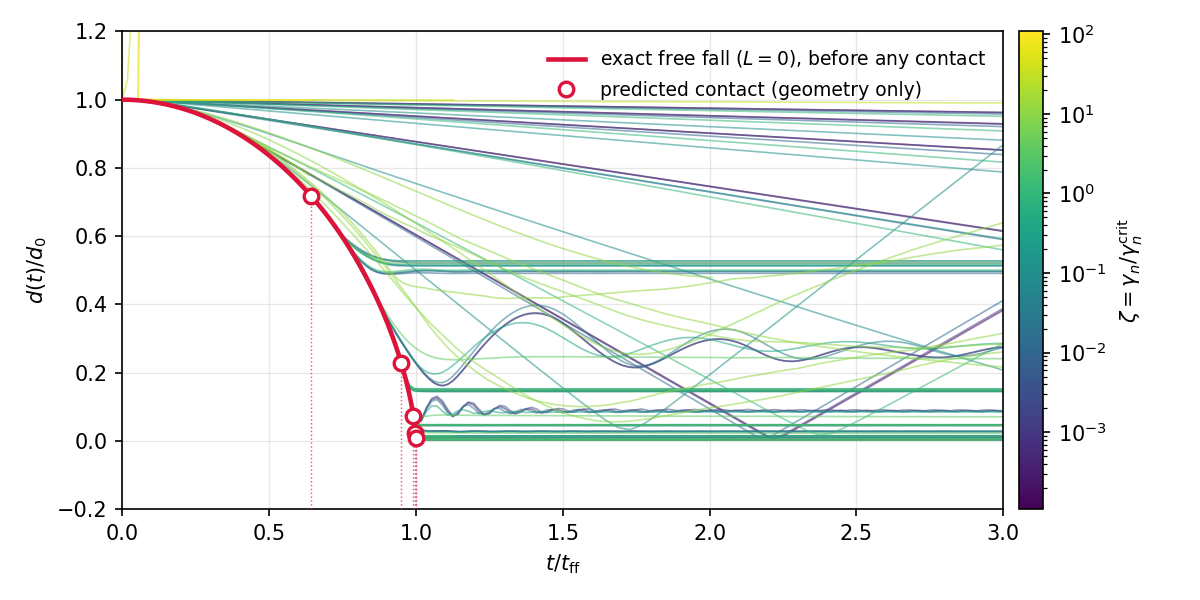}
\caption{Center-to-center separation $d(t)/d_0$ against rescaled time $t/t_{\text{ff}}$ for all $125$ runs of the sweep (thin curves, colored by $\zeta$), against the exact free-fall solution (thick red curve) and the geometric first-contact point for each sampled $d_0$ (open circles). Trajectories that remain flat near $d/d_0\approx1$ out to $\tau=3$ are the simulations whose fixed step budget (Sec \ref{sec:experiment:sweep}) expires during the initial infall, before first contact is ever reached.}
\label{fig:freefall_collapse}
\end{figure}

Figure \ref{fig:freefall_collapse} shows this parameter-free curve with the measured $d(t)/d_0$ trajectory of every one of the $125$ runs, colored by $\zeta$. Every trajectory collapses onto the analytic curve to within plotting accuracy from $t=0$ up to the geometrically predicted first-contact point $(\tau_c,x_c)$ (open circles, one per sampled $d_0$, computed from $d_{\text{contact}}/d_0$ alone), independent of $\rho$, $\gamma_n$, or the eventual outcome. This is a direct argument demonstrating that the model accurately reproduces two-body free fall before any contact occurs, and it determines the point at which the point-mass approximation must be abandoned in favor of full dynamics. Past that marked point, trajectories immediately fan out according to $\zeta$, as expected once contact damping starts to compete with the kinetic energy delivered by the fall.

\subsection{Outcome statistics and the reach of a fixed integration window}
\label{sec:results:outcomes}

On the $125$ runs, $71$ ($56.8\%$) are classified \textit{merged}, $42$ ($33.6\%$) \textit{infalling}, and $12$ ($9.6\%$) \textit{bouncing}. The purely geometric ratio $v_{\text{imp}}/v_{\text{esc}}=\sqrt{1-d_{\text{contact}}/d_0}$ takes exactly the five values dictated by the five sampled separations (Table \ref{tab:outcomes}), independent of density, since both velocities scale identically with $\sqrt{M_{\text{tot}}}$.

\begin{table}[tbh]
\caption{Outcome counts (of $25$ runs each) by initial separation $d_0$ (equivalently, by $v_{\text{imp}}/v_{\text{esc}}$, independent of density).}
\label{tab:outcomes}
\begin{ruledtabular}
\begin{tabular}{ccccc}
$d_0\,[\si{m}]$ & $v_{\text{imp}}/v_{\text{esc}}$ & merged & bouncing & infalling \\ \hline
$31.6$ & $0.532$ & $18$ & $7$ & $0$ \\
$100$ & $0.880$ & $18$ & $3$ & $4$ \\
$316$ & $0.964$ & $13$ & $1$ & $11$ \\
$1000$ & $0.989$ & $13$ & $1$ & $11$ \\
$3162$ & $0.996$ & $9$ & $0$ & $16$ \\
\end{tabular}
\end{ruledtabular}
\end{table}

This trend, however, is still not primarily a statement about reaccumulation dynamics: it remains, even at the enlarged step budget, overwhelmingly a statement about how much of the fixed-step-budget window (Sec \ref{sec:experiment:sweep}) is spent in the pre-contact free fall of Sec \ref{sec:results:freefall}. Of the $42$ \textit{infalling} runs, $40$ never reach a separation below half of $d_0$ at any point in their recorded history, and at the largest sampled separation, none of the $16$ still \textit{infalling} runs there ever comes within $\SI{25}{m}$ (essentially $d_{\text{contact}}$) of contact, while every one of the $9$ that do reach that band's merger is (unsurprisingly) counted as \textit{merged}, not \textit{infalling}. This is corroborated independently by the bounce count of Sec \ref{sec:results:bounces}: every single one of the $42$\textit{infalling} runs records exactly zero sign changes of the approach velocity. It is consistent with a run that never gets far enough to complete even its first contact, let alone bounce. 

Four runs in the entire grid (density $\SI{e5}{kg/m^3}$, $d_0=\SI{100}{m}$) are caught genuinely near, though not quite past, first contact ($46$--$54\,\si{m}$, against a contact distance of $\SI{22.7}{m}$) at the moment their window closes. These four belong to the density corner flagged as energetically unreliable below, so their proximity to contact should be read as geometrically indicative rather than quantitatively trustworthy. The \textit{infalling} label remains predominantly an artifact of a step budget sized from the point-mass free-fall estimate $t_{\text{ff}}(d_0)$ rather than from the substantially longer time needed to also resolve the ensuing contact sequence at the lowest sampled densities and the largest sampled separations.

A second, opposite effect operates at the single grid corner of highest density and closest separation ($\rho=\SI{e5}{kg/m^3}$, $d_0=\SI{31.6}{m}$): there, the same step-budget heuristic collapses to its floor of $200$ steps for all five sampled $\gamma_n$, because the point-mass estimate $4\,t_{\text{ff}}(d_0)$ is itself already shorter than that threshold. These five runs record the most violent encounters of the entire sweep on purely geometric grounds and are still separating, at up to $\SI{8.7}{m/s}$, when the run ends. However, it is precisely in this area of density that the code’s own energy balance loses its reliability. Therefore, from a quantitative standpoint, one should not attach too much importance to either the damage indicated or the rate at which it propagates.

\subsection{Representative final states}
\label{sec:results:gallery}

\begin{figure*}[tbh]
\centering
\includegraphics[width=0.78\textwidth]{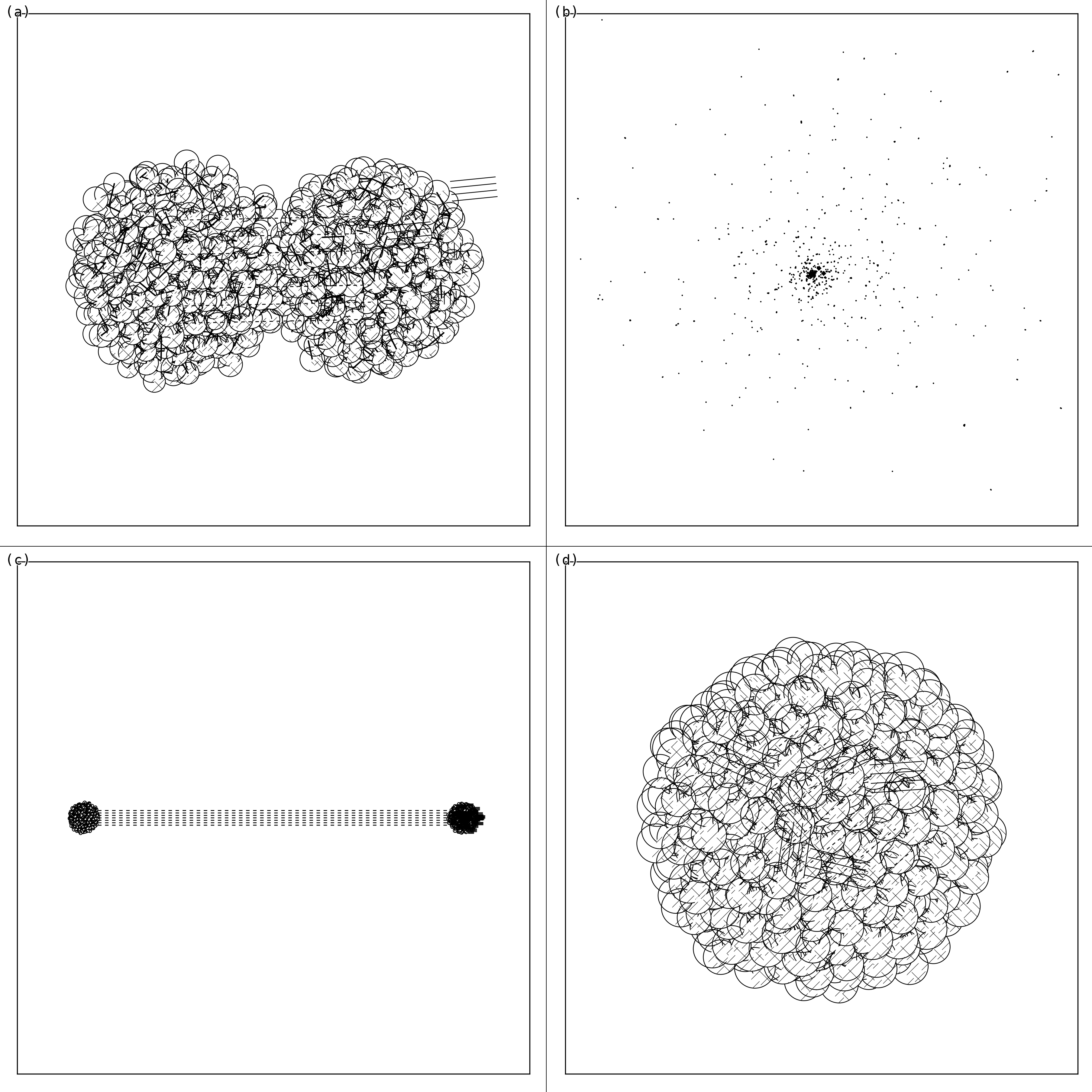}
\caption{Final particle-level state of four representative runs. (a) \textit{merged}, undamaged. (b) \textit{bouncing}, revealed at this step budget as an evaporating swarm rather than two intact bodies. (c) \textit{infalling}, fully separate and undamaged, with the mutual gravitational field bundle spanning the gap. (d) \textit{merged} but completely debonded, with fast-moving grains marked by directional strokes. Panel (b-c) is shown at a wider physical scale than the others, so individual grains fall below one pixel and appear as points rather than resolved, hatched circles.}
\label{fig:gallery}
\end{figure*}

Fig \ref{fig:gallery} shows the full particle-level final state of four representative runs, one per broad outcome, rendered directly from each run's own final particle positions, radii, body labels, and intact-bond list. Grain shading is produced exclusively by angularly-clipped cross-hatching under a single fixed light direction, never by a continuous tone. Intact cohesive bonds are drawn as solid links; a handful of the fastest-moving grains are marked by short directional strokes rather than by an instantaneous velocity field. And the mutual gravitational field of Eq \ref{eq:gravity} is superposed as a bundle of regularly dashed lines connecting the two bodies' centers of mass, drawn without regard to occlusion by the grains themselves.

Fig \ref{fig:gallery}.a) ($\rho=\SI{10}{kg/m^3}$, $d_0=\SI{31.6}{m}$, $\gamma_n=\SI{100}{N.s/m}$) is \textit{merged}: the two composites have fallen into, and remain in a single aggregate. Fig \ref{fig:gallery}.c) (same $\rho$ and $\gamma_n$ but $d_0=\SI{316}{m}$) is \textit{infalling}: by the end of the run the separation has fallen from $316$ to $\SI{229}{m}$, but the two bodies remain fully apart and intact. Fig \ref{fig:gallery}.d) ($\rho=\SI{e4}{kg/m^3}$, $d_0=\SI{3162}{m}$, $\gamma_n=\SI{e6}{N.s/m}$) is \textit{merged} as well ($d_{\text{final}} \approx\SI{6}{m}$) but with $D_{\text{final}}=1$ ($D$ is described in Sec \ref{sec:results:damage}): no bond survives, and the directional strokes mark a population of fast-moving grains within the reassembled pile consistent with a large kinetic energy. Fig \ref{fig:gallery}.b) ($\rho=\SI{e3}{kg/m^3}$, $d_0=\SI{100}{m}$, $\gamma_n=\SI{e6}{N.s/m}$) is \textit{bouncing}, but at the resolution of individual grains this label does not mean two intact bodies caught mid-separation: extending the integration this far instead reveals a slowly evaporating swarm, a minority of grains still clustered near a common center while the rest have individually drifted, under self-gravity alone, out to distances far beyond the pair's own nominal radius. This is a direct particle-level illustration of the sub-percolation-threshold bond network: with a mean bond count below one per grain, a sizable fraction of each composite was never cohesively attached to anything in the first place, and is consequently free to disperse under even the modest gravity of this toy system once given the several-\si{ks} window this run's step budget now allows.

\subsection{Damage and fragmentation as a function of impact stress}
\label{sec:results:damage}

For the subset of simulations that do reach contact, the relevant impact severity is not $v_{\text{imp}}$ alone but its ratio to the bond network's own force scale. An unbonded grain-grain contact closing at speed $v_{\text{imp}}$ loads a linear contact spring of stiffness $k_n$ and reduced mass $m_{\text{eff}}$ up to a peak elastic force $v_{\text{imp}}\sqrt{k_n m_{\text{eff}}}$ with $m_{\text{eff}}=m_i m_j/(m_i+m_j)$ the reduced mass of the two colliding grains.

Comparing this force scale to the bond-breaking force $\sigma_c A$ of Eq \ref{eq:bond_stress} (where $A=\pi\bar r^2$) defines the dimensionless impact-stress number
\begin{equation}
\Pi_\sigma = \frac{v_{\text{imp}}\sqrt{k_n m_{\text{eff}}}}{\sigma_c A}
\label{eq:pi_sigma}
\end{equation}
which spans four decades over the study, from $1.9\times10^{-3}$ to $35.9$. However, as established in section \ref{sec:results:outcomes}, all simulations greater than $\Pi_\sigma\approx3.6$ belong to a single sampling density whose energy balance is unreliable. Consequently, the reliable subset of $90$ simulations only ever samples values of $\Pi_\sigma$ up to $3.59$.

\begin{figure}[tbh]
\centering
\includegraphics[width=\linewidth]{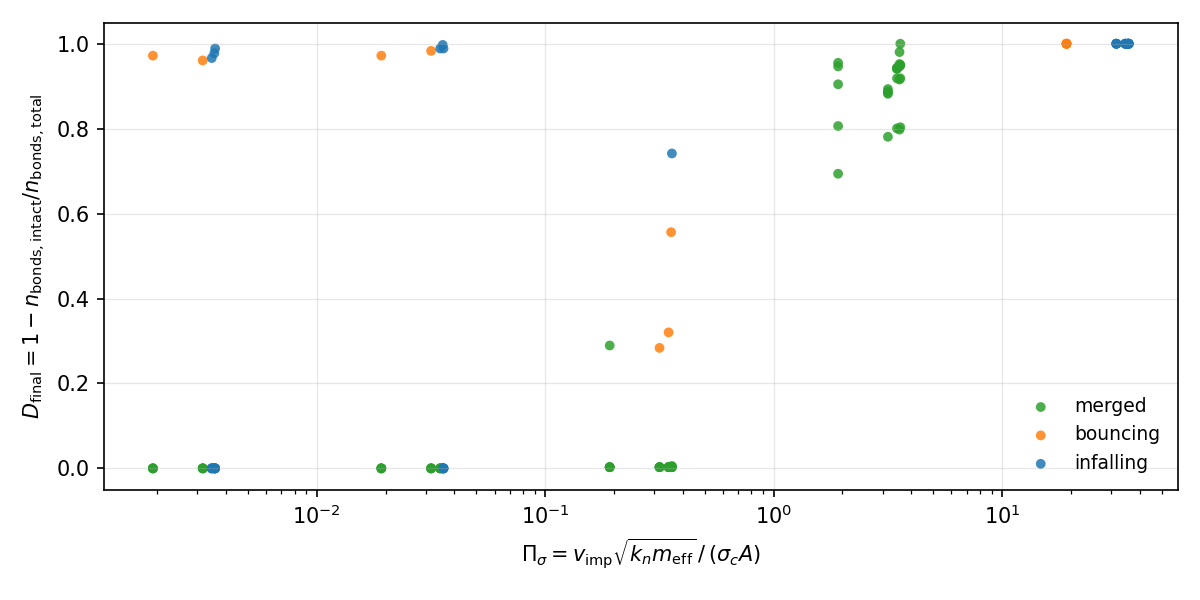}\\[4pt]
\includegraphics[width=\linewidth]{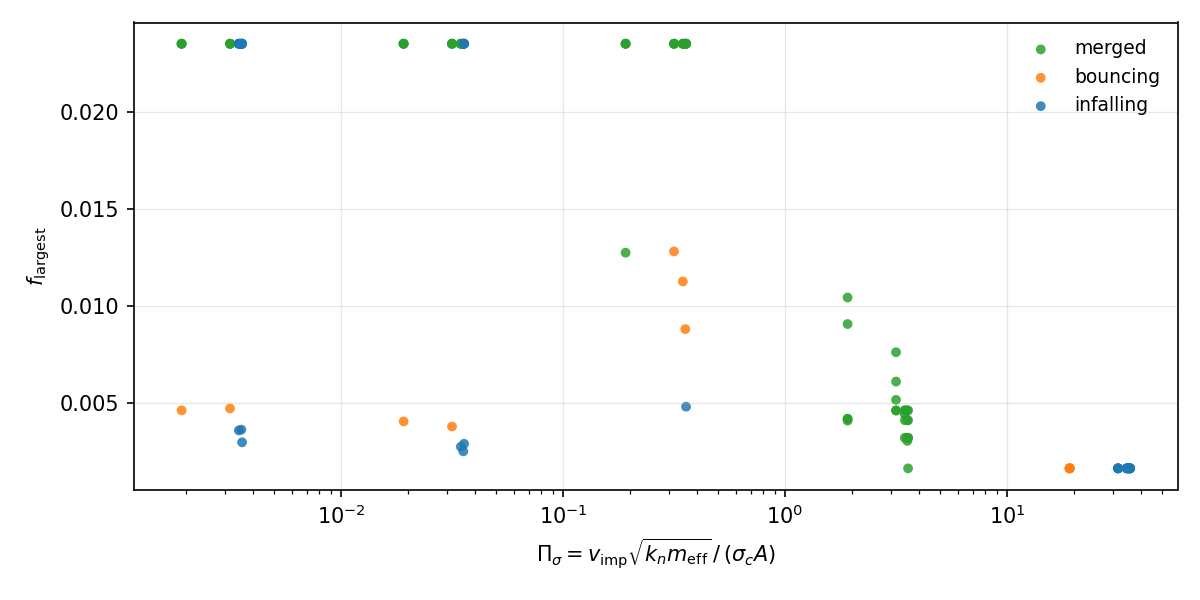}
\caption{Final damage $D_{\text{final}}$ (top) and largest-fragment mass fraction $f_{\text{largest}}$ (bottom) against the dimensionless impact-stress number $\Pi_\sigma$ of Eq \ref{eq:pi_sigma}, for all $125$ simulations.}
\label{fig:damage_fragmentation}
\end{figure}

Figure \ref{fig:damage_fragmentation} plots the final damage $D_{\text{final}}$ and largest-fragment mass fraction $f_{\text{largest}}$ of every run against $\Pi_\sigma$, including the $35$ excluded ones (identifiable by their label). The damage variable, determined such as:
\begin{equation}
D = 1 - \frac{N_{\text{intact}}}{N_{\text{total}}}
\end{equation}
tracks the fraction of the initial cohesive network that has failed.

\begin{figure}[tbh]
\centering
\begin{tikzpicture}[scale=0.62,every node/.style={font=\small}]
  \definecolor{grainA}{RGB}{176,196,222}
  \node[font=\bfseries] at (2,4.6) {intact network};
  \foreach \i/\x/\y in {1/0/2.6, 2/1.2/3.6, 3/2.6/3.9, 4/3.9/3.2,
                        5/4.0/1.8, 6/2.8/0.9, 7/1.3/1.1} {
     \fill[grainA] (\x,\y) circle (0.42);
     \draw (\x,\y) circle (0.42);
  }
  \draw[thick] (0,2.6)--(1.2,3.6) (1.2,3.6)--(2.6,3.9) (2.6,3.9)--(3.9,3.2)
        (3.9,3.2)--(4.0,1.8) (4.0,1.8)--(2.8,0.9) (2.8,0.9)--(1.3,1.1)
        (1.3,1.1)--(0,2.6) (2.6,3.9)--(2.8,0.9);
  \node[red,font=\bfseries] at (2.7,2.35) {$\times$};

  \node[font=\bfseries] at (2,0.0) {two fragments after breakage};
  \foreach \i/\x/\y in {1/0/-2.6, 2/1.2/-1.6, 3/2.6/-1.3} {
     \fill[grainA] (\x,\y) circle (0.42);
     \draw (\x,\y) circle (0.42);
  }
  \foreach \i/\x/\y in {4/3.9/-2.0, 5/4.0/-3.4, 6/2.8/-4.3, 7/1.3/-4.1} {
     \fill[orange!35] (\x,\y) circle (0.42);
     \draw (\x,\y) circle (0.42);
  }
  \draw[thick] (0,-2.6)--(1.2,-1.6) (1.2,-1.6)--(2.6,-1.3);
  \draw[thick] (3.9,-2.0)--(4.0,-3.4) (4.0,-3.4)--(2.8,-4.3) (2.8,-4.3)--(1.3,-4.1);
  \draw[thick,gray,dashed] (2.6,-1.3) -- (3.9,-2.0);
  \node[red,font=\scriptsize,anchor=west] at (4.15,-1.35) {broken};
\end{tikzpicture}
\caption{Fragmentation as connected components of the intact-bond graph. The failure of a single bond (top, marked $\times$) can split one connected cluster of seven grains into two fragments of three and four grains respectively (bottom); the largest-fragment mass fraction $f_{\text{largest}}$ is computed from the union-find decomposition of this graph at every measure step.}
\label{fig:fragments}
\end{figure}
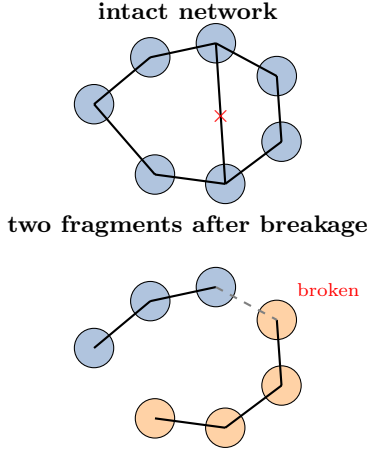

To define $f_{\text{largest}}$, I define a fragment as a connected component of a graph whose vertices are the grains and whose edges are the bonds still intact at that instant. Two grains merely touching, without an intact bond between them, are \emph{not} considered part of the same fragment, since an unbonded contact can slip apart at any subsequent step. Connected components are extracted with a union-find (disjoint-set) structure with path compression, run over the current set of intact bonds, and the mass fraction carried by the single largest component,
\begin{equation}
f_{\text{largest}} = \frac{\max_c M_c}{\sum_c M_c} ,
\end{equation}
is reported alongside the total fragment count $N_{\text{frag}}$. This mass fraction of the largest fragment is shown schematically in Figure \ref{fig:fragments}.

The interpretation below applies only to the $90$ retained simulations. No damage is recorded below $\Pi_\sigma\approx0.036$. $D_{\text{final}}$ becomes systematically positive by $\Pi_\sigma\approx0.19$ and within this trustworthy subset, already reaches complete failure ($D_{\text{final}}=1$, $f_{\text{largest}}=0.0016$, one surviving grain) by $\Pi_\sigma\approx3.59$. The transition to a complete breakage therefore lies within (and not beyond) the explored range. The collapse in $\Pi_\sigma$ is not exact: at fixed $\Pi_\sigma\approx1.92$, for instance, $D_{\text{final}}$ still ranges from $0.69$ to $0.96$ across the five sampled $\gamma_n$. The residual dispersion is attributable to $\zeta$, which determines the number of loading cycles (Section \ref{sec:results:bounces}) to which the surviving network is subjected before the end of the test, rather than to $\Pi_\sigma$, which incorrectly measures the severity of a single impact. $\Pi_\sigma$ therefore determines the overall extent of damage and fragmentation, but does not by itself determine the exact value in a given simulation.

\subsection{Restitution and the number of bounces before capture}
\label{sec:results:bounces}

\begin{figure}[tbh]
\centering
\includegraphics[width=\linewidth]{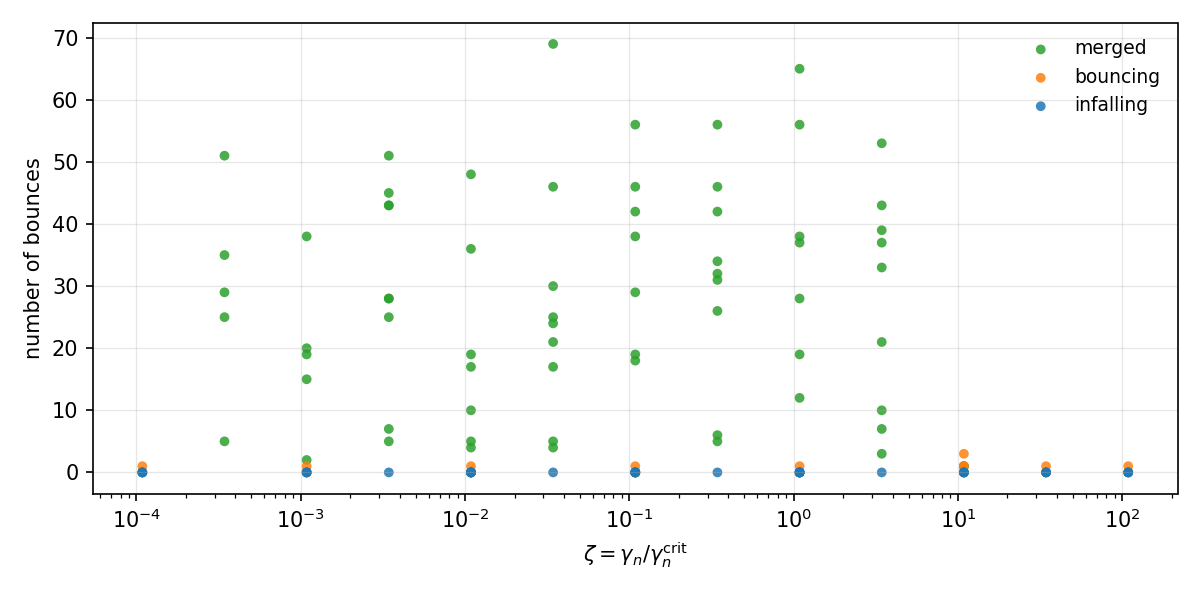}
\caption{Number of sign changes of the radial approach velocity recorded over a run, against the dimensionless damping ratio $\zeta=\gamma_n/\gamma_n^{\text{crit}}$. Every \textit{infalling} simulation records exactly zero bounces.}
\label{fig:bounces}
\end{figure}

Figure \ref{fig:bounces} reports the number of sign changes of the radial approach velocity recorded before the end of each simulation as a function of $\zeta$, for the $90$ simulations unaffected by either pathology of Sec \ref{sec:results:outcomes}. Here, $\zeta\equiv\gamma_n/\gamma_n^{\text{crit}}$ denotes the dimensionless contact damping ratio. Two regimes stand out. For $\zeta\gtrsim3$ ($20$ runs), only $60\%$ merge (or are caught mid-fall) within zero to three bounces. The others require between $4$ and $53$ bounces despite their high damping ratio. A single sufficiently overdamped contact removing most of the kinetic energy delivered by the fall, consistent with the analytic single-contact restitution coefficient $e(\zeta)=\exp(-\pi\zeta/\sqrt{1-\zeta^2})$ falling rapidly toward $0$ over the same range, is therefore at best a partial description even at $\zeta\gtrsim3$, and the qualitative claim that high damping alone guarantees fast capture does not survive the longer integration. For $\zeta\lesssim1$, merged runs require anywhere from a handful to as many as $69$ bounces, with substantial scatter at fixed $\zeta$ rather than a single well-defined count, before enough energy has been removed for the pair to settle. This remains the qualitative behavior expected once each bounce only removes a fraction $1-e(\zeta)^2\ll1$ of the orbital kinetic energy, but, as before, no quantitative fit to $e(\zeta)$ is attempted, since $e(\zeta)$ is a single-grain-contact quantity, whereas the number of bounces of a two-body, many-grain composite additionally depends on the evolving bond damage (Sec \ref{sec:results:damage}). The three \textit{bouncing} simulations not otherwise flagged in Sec \ref{sec:results:outcomes} sit at $\zeta\approx10.9$, with only one to three bounces recorded and a small residual outward velocity ($<\SI{0.2}{m/s}$) at the end of the run: these remain consistent with a merger simply not yet complete within the allotted window, rather than with either censoring mechanism identified above.

\section{Conclusion}
\label{sec:conclusion}

I presented a discrete-element model combining flexible-sphere contact, cohesive connections, an irreversible failure criterion, and direct pair-wise self-gravity, along with the corresponding numerical protocol. I built stochastic aggregate generation, a controlled two-body encounter, and a systematic reaccumulation sweep. 

This study presents three findings. First, the model reproduces the exact, parameter-free solution for two-body free fall prior to the first contact for every combination of density, distance, and damping examined. Next, once expressed in terms of the dimensionless impact-stress number $\Pi_\sigma$ from equation \ref{eq:pi_sigma}, the response of the connection network in terms of damage and fragmentation clearly splits into an undamaged regime ($\Pi_\sigma\lesssim0.036$), a partially damaged regime with residual dispersion determined by the damping ratio $\zeta$, and complete failure starting at $\Pi_\sigma\approx3.59$. Finally, the number of low-speed bounces required before two bodies settle into a single remnant is highly variable at fixed damping ratio $\zeta$, showing that the single-contact restitution coefficient $e(\zeta)$ is at best a partial predictor of capture time.

This study has some limitations. Each point of the parameter study relies on a single stochastic packing realization, so the reported scatter cannot be separated from packing-to-packing variability. Grains are not allowed to rotate, so contact and bond torques are absent from the model. Self-gravity is evaluated by direct $O(N^2)$ summation, limiting the particle counts that can be reached. Future work should average over several realizations per grid point, add rotational degrees of freedom, and explore a tree-based gravity solver to reach more realistic grain numbers.

\begin{acknowledgments}
The author thanks the open-source scientific Python and C++ ecosystems on which this work relies.
\end{acknowledgments}

\bibliography{references}

\end{document}